\documentclass[11pt]{article}
\usepackage{amssymb, amsmath}
\usepackage{mathtools}
\def\be{\begin{equation}}
\def\ee{\end{equation}}
\def\ba{\begin{eqnarray}}
\def\ea{\end{eqnarray}}
\def\nn{\nonumber}
\def\lb{\label}
\def\dfrac{\displaystyle\frac}
\def\bb{\bibitem}
\def\E{{\cal E}}
\def\tJ{\tilde{J}}
\renewcommand{\theequation}{\arabic{section}.\arabic{equation}}
\begin{document}
\begin{titlepage}
\date{}
\title{\begin{flushright}\begin{small}    LAPTH-043/19
\end{small} \end{flushright} \vspace{1.5cm}
On the Smarr formulas for electrovac spacetimes with line singularities  }
\author{G\'erard Cl\'ement$^a$\thanks{Email: gclement@lapth.cnrs.fr},
Dmitri Gal'tsov$^{b,c}$\thanks{Email: galtsov@phys.msu.ru} \\ \\
$^a$ {\small LAPTh, Universit\'e Savoie Mont Blanc, CNRS, 9 chemin de Bellevue,} \\
{\small BP 110, F-74941 Annecy-le-Vieux cedex, France} \\
$^b$ {\small Department of Theoretical Physics, Faculty of Physics,}\\
{\small Moscow State University, 119899, Moscow, Russia }\\
$^c$ {\small  Kazan Federal University, 420008 Kazan, Russia}}

\maketitle

\begin{abstract}
Using the revised Komar-Tomimatsu approach, we derive Smarr mass formulas for
stationary axisymmetric solutions of the Einstein-Maxwell equations containing line singularities (defects) on the polar axis. In terms of the rod structure associated with Weyl representation of the metric, the horizons and  the defects are formally similar up to differences due to their timelike/spacelike character. We derive (previously unknown or incorrect) horizon and global Smarr formulas in presence of a Newman-Unti-Tamburino (NUT) parameter.
To   avoid the  divergence of the Komar angular momentum of semi-infinite Dirac and Misner strings,  it is necessary to use a symmetric tuning.  We also note that the horizon mass Smarr formula does not include either magnetic charge, or  NUT parameter, correcting some statements in the literature. The contribution of each Misner string to the total mass consists in an angular momentum term, an electric charge term, and a length term, which can also be presented as the product of the spacelike analogue of surface gravity and the area of the string.
\end{abstract}
\end{titlepage}
\setcounter{page}{2}

\setcounter{equation}{0}
\section{Introduction}

Smarr's original formula \cite{smarr,carter} relates the mass, the angular momentum, the horizon area and
the electric charge of regular asymptotically flat black holes in Einstein-Maxwell theory.
Its differential form was recognized as the first of the four laws of  black hole mechanics
\cite{Bardeen:1973gs} and then interpreted as a genuine  first law of thermodynamics \cite{Hawking:1976de}.
The rationale for this interpretation is the existence of a regular Euclidean continuation (instanton)
of the black hole solution and the identification of the on-shell instanton action with a suitable
thermodynamic potential \cite{Gibbons:1976ue}. The first law was generalized to asymptotically flat black
holes in a number of theories including supergravity, higher curvature gravity, as well as asymptotically
AdS solutions and solutions in higher dimensions. Meanwile, numerous attempts in the past
\cite{Hunter:1998qe,Hawking:1998jf, Carlip:1999cy, Mann:1999pc, Mann:1999bt, Astefanesei:2004ji,Mann:2004mi, Ghezelbash:2007kw} to find a consistent generalization of the first law in presence of a NUT parameter, or magnetic mass, have not yet been fully successful.
One of the reasons is that to avoid a line singularity (Misner string) on the polar axis, it was common
to impose the time periodicity condition suggested by Misner \cite{Misner:1963fr}. But this takes the
Lorentzian solution  beyond the physically meaningful class of solutions. It is also worth noting, that in
the instanton sector the Misner and Hawking periodicities can not be consistently imposed together in the
rotating case \cite{Lapedes:1980qw,Chen:2010zu}.

To attempt to give physical meaning to the solutions with NUTs, an alternative interpretation of the Misner
string due to Bonnor  \cite{bonnor,Sut} can be  invoked. In this interpretation, similarly to the infinitely
thin, semi-infinite solenoid which generates the Dirac magnetic monopole, the Misner string is a truly physical
(singular) material object: a distribution-valued gravimagnetic flux. We have shown that such Misner strings,
contrary to previous statements in the literature, are transparent to geodesic motion and that, while they are
surrounded by closed timelike curves, these cannot be geodesic
\cite{Clement:2015cxa,Clement:2015aka,Clement:2016mll}. Adopting this interpretation, a thermodynamics for
NUTty black holes was suggested recently in a series of papers
\cite{Kubiznak:2019yiu,Ballon:2019uha,Bordo:2019tyh,Bordo:2019rhu}.

Meanwhile, in our opinion, the interpretation of the NUT terms in the mass formulas still leaves questions.
Actually, the derivation of Smarr relations in the past was done only for regular spacetimes. Solutions with
Dirac and Misner strings are beyond this class, belonging rather to spaces with line singularities
\cite{Israel:1976vc}. Recently, the Smarr mass formulas were rederived in the case when the magnetic charge
is present, with an associated Dirac string \cite{Clement:2017otx}. Using the revised Tomimatsu
\cite{tom83,tom84} version of the Komar \cite{komar} integrals for stationary axisymmetric spacetimes,
we demonstrated that, in presence of gravity, the Dirac string is heavy and contributes to the total mass
of the dyonic Reissner-Nordstr\"om black hole. At  the same time, the magnetic term was shown not to enter
into the Smarr  formula for the horizon mass\footnote{It is worth noting that the preprint \cite{manko2015}
criticized by us in \cite{Clement:2017otx} was nevertheless published later under a different title
\cite{manko2017} with the same erroneous statement.}. In fact, in more general situations with several
disconnected black holes joined by struts and Misner and Dirac strings, typical for various double black hole
solutions, the same kind of calculations proved to be successful
\cite{Clement:2017kas,Clement:2018nmt,Clement:2019clb}.

Here we extend this approach to the general class of line singularities, using for their description  the rod structure formalism \cite{Harmark:2004rm, Chen:2010zu} developed for spacetimes with two commuting Killing vectors ($D-2$ in the higher-dimensional case). Generically, the rods represent both black hole horizons and line defects on the polar axis on an equal footing, so it turns out that Smarr formulas can be obtained for all of them in the same way. Applying this formalism to rotating black holes with NUTs, one can give a clear physical interpretation of some problematic terms encountered in previous proposals for NUT thermodynamics, including the recent ones  \cite{Kubiznak:2019yiu,Ballon:2019uha,Bordo:2019tyh,Bordo:2019rhu}.

\setcounter{equation}{0}
\section{Rod structure}
Stationary axially symmetric solutions of the Einstein-Maxwell equations generically contain line singularities on the polar axis. A general classification of such solutions can be done in terms of the rod structure approach \cite{Harmark:2004rm}. In Weyl coordinates $x^a, \rho, z$, where $ x^a=t,\,\varphi$ one can present  the line element as
 \begin{equation}                    \label{Gzr}
 ds^2 = G_{ab}(\rho, z) dx^a dx^b + e^{2\nu}(d\rho^2 +  dz^2)\,,
\end{equation}
where the Gram matrix $G_{ab}$ and $\nu$ are functions of $(\rho, z)$, and the coordinate $\rho$ is related to  $G$ via
\begin{equation}                     \label{rho}
 \rho = \sqrt{|\det G|}\,.
\end{equation}
Any solution of the above class can be uniquely specified by  the matrix $G$ at the polar axis $\rho=0$
\cite{Harmark:2004rm}, this applies also to D-dimensional metrics with $D-2$ commuting Killing vectors.
From (\ref{rho}), it is clear that the Gram matrix is non-degenerate as long as $\rho >0$. At $\rho=0$,
it  degenerates, so the kernel of the boundary matrix $G(\rho=0, z)$ becomes nontrivial, i.e.,
$\dim \ker G(0, z) \ge 1$. It can be proved that, if the kernel has dimension higher than one, there will
be a strong curvature singularity on the axis~\cite{Harmark:2004rm}. If $\dim \ker G(0, z) = 1$ exactly,
except for a finite number of isolated points  $z_n,\; n=1, \dots, N$, one encounters only weak
distributional singularities on the polar axis, or no singularities at all.  The above isolated points,
called turning points, will be ordered as $z_1 < z_2 < \dots < z_N$. The set $z_n$ divides the polar axis
$z$ in $N+1$ intervals $(-\infty, z_1], [z_1, z_2], \dots, [z_N, +\infty)$ which are called rods (we will
label two semi-infinite rods by $n=\pm$, and the remaining finite ones by an index $n$ corresponding to the
left bound of the interval).

For each  rod one defines the eigenvectors $l_n  \in
 \mathbb{R}^2$, belonging to the kernel of  $G(0,z)$:
\begin{equation}
 G_{ab}(0, z)l_{n}^a = 0, \quad z  \in [z_n, z_{n+1}]\,.
\end{equation}
An important property of the eigenvectors is that they are  {\em constant} along each rod
\cite{Harmark:2004rm}, this will be essentially used in the following.
The direction of a rod defines a Killing vector field $l_n^\mu$ of the space-time, written
in the basis consisting of  $k=\partial_t$ and $m=\partial_\varphi$. Along a specific rod the associated
Killing vector field vanishes. Near the interior of the rod $\rho\to 0,\; z_n<z<z_{n+1}$ the norm
$l_{n}^2=G_{ab}l_{n}^a l_{n}^b\sim \pm a(z) \rho^2$ and $e^{2\nu}\sim c^2 a(z)$ ($c$ constant) in the leading
order in $\rho$. Thus the quantity $\rho^{-2} e^{-2\nu}l^2 $ has a finite limit on the polar axis and will be constant along the corresponding rod.
In Lorentzian spacetime this quantity can be negative, positive or zero, the rod is said to be time-like, space-like or light-like, respectively.
 We will be interested in the solutions characterized by timelike and spacelike rods.
The latter potentially correspond to  line singularities \cite{Israel:1976vc},  the finite timelike rods --- to the Killing horizon  of $k=\partial_t+\Omega_H\partial_\varphi$,  rotating with the angular velocity $\Omega_H$. The associated surface gravity is
\be\lb{karod}
\kappa_H=\left(-l_{\mu ;\nu} l^{\mu ;\nu}/2\right)^{1/2}=\lim_{\rho\to 0} \left(- \rho^{-2}e^{-2\nu}G_{ab}l_{H}^a l_{H}^b  \right)^{1/2},
\ee
 By the well-known theorems of black hole theory, both $\kappa_H$ and $ \Omega_H$ are constant on the horizon, consistently with the constancy of the directional vectors along the horizon rod proved in \cite{Harmark:2004rm}. Multi-black hole solutions may have several finite timelike (horizon) rods.
Timelike rods of infinite length correspond to  acceleration horizons (not considered here).

For spacelike rods one finds that similar limits define potential conical defects, which are avoided if the coordinate $\eta$ associated with the Killing vector field $l=\partial_\eta$ is identified with the period
\be\label{ident}
\Delta \eta=2\pi\lim_{\rho\to 0} \left(\rho^{2}e^{2\nu}\left(G_{ab}l_{n}^a l_{n}^b  \right)^{-1}\right)^{1/2}.
\ee
In what follows this condition  {\em is not}  assumed, i.e. we will deal with true line singularities.

For some purposes, it is convenient to normalize the rod directional vectors so that the  square roots in
(\ref{karod}), (\ref{ident}) be equal to one \cite{Chen:2010zu}. The normalized direction of the horizon rod then will be
\be\label{lh}
l_H=\left( 1/\kappa_H,\;  \Omega_H/\kappa_H\right).
\ee
The normalized spacelike rod directional vectors can be presented in the same form
\be\label{ln}
l_n=\left( 1/\kappa_n,\;  \Omega_n/\kappa_n\right),
\ee
where the parameter $\kappa_n$ is
\be\lb{kadef}
\kappa_n= \lim_{\rho\to 0} \left( \rho^{-2}e^{-2\nu}G_{ab}l_{n}^a l_{n}^b  \right)^{1/2}.
\ee
With the normalized spacelike directional vector, the period (\ref{ident}) will be $2\pi$.
The spacelike rod directions  define the Killing vector field in spacetime, which is {\em spacelike} outside the rod but becomes null on it, therefore (\ref{kadef}) can be called  spacelike ``surface gravity''\footnote{Apparently the spacelike nature of the Killing corotating vector in the vicinity of the Misner string was not noticed in \cite{Bordo:2019rhu}:  the sign in the definition of  $\kappa^2$ after the formula (7) in  \cite{Bordo:2019rhu}
is wrong.}. The Killing horizon of the spacelike corotating Killing vector may seem similar to the internal Cauchy horizon in the interior of a black hole, but it is not. In the latter case the horizon separates timelike and spacelike regions, while in the string case there is no internal regions where this vector becomes timelike again.

A well-known example of a line singularity is an infinitely thin cosmic string; the corresponding distributional energy-momentum tensor has a one-dimensional equation of state $ \epsilon = -p $. The  global four-dimensional metric contains conical singularities along the string. Such singularities are weak, they can be smoothed out by a suitable matter source, leading to a ``thick'' model of the cosmic sting. In this case, we are accustomed to recognize the line singularity as physical.
 One often encounters such cosmic string singularities with $\epsilon<0$ (struts) in binary black hole solutions. 

The weak line singularities which we shall discuss here are the Misner strings in the Taub-NUT and other spacetimes endowed with NUT parameter. Although the corresponding distributional stress-energy tensor does not have such a simple physical interpretation, mathematically it belongs to the same class. Double black holes  \cite{Clement:2017kas,Clement:2018nmt,Clement:2018nmt,Clement:2019clb} carrying magnetic and NUT charges may contain struts, ensuring their equilibrium, with the properties of cosmic strings with positive or negative conical defect angles, as well as  Dirac and Misner string carrying distributional magnetic and gravimagnetic fields, respectively. We will be not interested in the detailed structure of the distributional  sources, but we would like to determine the masses, angular momenta, and electric charges generated by them. These are most naturally defined by the Komar conserved charges \cite{komar} considered in the next section.
Since the constancy of the directional vectors on the rods is also valid for spacelike rods, one can define the mass, angular momentum and electric charge of the  line singularities, in the same way as for the black hole horizons, in terms of the rod directional parameters $\Omega_n,\,\kappa_n$.

We illustrate the above construction for the Kerr-NUT solution. In the Boyer-Lindquist coordinates the metric is
\begin{align}\lb{s}
&ds^2=-\frac{\Delta}{\Sigma}\left(dt+P_\theta d\varphi\right)^2+\Sigma\left(\frac{dr^2}{\Delta}+d\theta^2\right)+\frac{\sin^2\theta}{\Sigma}\left(adt-P_r d\varphi\right)^2,\\
&P_\theta=2n\cos\theta+2s-a\sin^2\theta,\quad P_r=r^2+a^2+n^2-2as, \\
& \Sigma = P_r + aP_\theta = r^2 + (n+a\cos\theta)^2, \quad \Delta=r^2-2mr+a^2-n^2,
\end{align}
where $m$, $n$ and $a$ are the mass, NUT and rotational parameters, and  the ``large gauge'' parameter $s$ regulates the relative strength of the two semi-infinite Mismer strings $\theta=0,\pi$. The corresponding Weyl coordinates are $\rho=\sqrt{\Delta}\sin\theta,\, z=(r-m)\cos\theta$.
In this case the rod structure consists of three rods $( -\infty,-z_H], [-z_H, z_H],   [z_H, \infty)$ with $z_H=r_H-m,\; r_H=m+\sqrt{m^2+n^2-a^2}$, joining pairwise at $z=\pm z_H$ and the directions (\ref{lh}, \ref{ln})
with the parameters
\be\lb{parakn}
 \kappa_H=\frac{r_H-m}{(mr_H+n^2-as)},\quad \Omega_H=\frac{a \kappa_H}{2(r_H-m)}, \quad \kappa_{\pm}=\frac1{2(n\pm s)},\quad\Omega_{\pm}= \mp\kappa_\pm.
\ee
The rod $l_H$ defines the spacetime Killing vector $\xi_H=\partial_t+\Omega_H\partial_\varphi$
which is timelike outside the horizon, and becomes null on it. The rods $l_\pm$ define the Killing vectors $\xi_\pm=\partial_t+\Omega_\pm\partial_\varphi$ which are {\em spacelike} outside the polar axis for $|z|>z_H$ and  become  null on the Misner strings. Their norm in the vicinity of the polar axis for $s=0$ is
\be
\xi_\pm^2=\frac{r^2 + (n\pm a)^2}{4n^2}\sin^2\theta.
\ee
The associated ``surface gravity'' $\kappa_\pm$, therefore, is not associated with particle acceleration, neither with a redshift factor, so it can hardly be interpreted as Hawking temperature.

In Euclidean signature metrics, in particular, in the Wick-rotated stationary axisymmetric metrics, all the rods
are spacelike. To justify thermodynamical interpretation of the differential Smarr formula, one has to calculate
the on-shell action of the  Euclidean instanton. A sufficient condition to ensure finiteness of the action is
regularity of the instanton solution. This means that the rod structure must be consistent with $2\pi$
periodicity rules for {\em all} rods. As the two neighboring rod directional vectors meet at the turning points,
one has  to make there an identification of two spacelike directions. This always can be done at one turning
point, but in the case of several such points, the corresponding pairs of  vectors  must be related  by
$GL(2,Z)$ transformations, otherwise one will have {\em orbifold} singularities \cite{Chen:2010zu}.
Unfortunately, this happens already for Kerr-NUT with a symmetrical tuning of the Misner strings
\cite{Chen:2010zu}  (in the Appendix we generalize the proof to the asymmetric case) and more general type D
instantons \cite{Lapedes:1980qw}.


\setcounter{equation}{0}
\section{Komar charges in presence of line singularities}
A convenient setting for calculating Komar integrals in spaces with weak line singularities on the axis is a scheme developed by Tomimatsu \cite{tom83,tom84} and corrected in
\cite{Clement:2017otx}. Here we reformulate the approach of \cite{Clement:2017otx} in terms of the rod structure.

Let an asymptotically locally flat spacetime have a regular event horizon $H$, possibly consisting of several disconnected pieces $H_i$ (multi-black holes), and a certain number of defects, both being represented by rods on the axis $\rho=0$ in Weyl coordinates. It is necessary to surround all the rods $n$ -- timelike (horizons) or spacelike (defects) -- by small cylinders  $\Sigma_n$. The physical domain (bulk) will thus be bounded by these cylinders and a two-dimensional sphere at infinity $\Sigma_\infty$.  Typical examples are  solutions with NUTs including single and binary black holes   \cite{Clement:2017kas,Clement:2018nmt,Clement:2019clb}.

The total Komar mass, angular momentum and electric charge of a stationary axisymmetric configuration
are given by the integrals over $\Sigma_\infty$:
 \be
M = \frac1{4\pi}\oint_{\Sigma_\infty}D^\nu k^{\mu}d\Sigma_{\mu\nu}, \quad
J = -\frac1{8\pi}\oint_{\Sigma_\infty}D^\nu m^{\mu}d\Sigma_{\mu\nu},\quad
 Q=   \frac1{4\pi}\oint_{\Sigma_\infty} F^{\mu\nu} d\Sigma_{\mu\nu}
\lb{koMJ}
 \ee
where $k^\mu = \delta^\mu_t$ and $m^\mu = \delta^\mu_\varphi$ are
the Killing vectors associated with time translations and rotations
around the $z$-axis, $D^\nu$ is the covariant derivative and $F^{\mu\nu}$ is the Maxwell tensor. By the Gauss theorem, the total electric charge is equal to the sum of the fluxes
through the various cylinders $\Sigma_n$:
 \be\lb{Qtot}
Q = \sum_n Q_n, \qquad Q_n = \frac1{4\pi}\oint_{\Sigma_n}F^{ti}d\Sigma_i.
 \ee
Consider now the Komar mass, which coincides with the ADM mass.
Because the integrand $D^\nu k^{\mu}$ is antisymmetric, one can
apply the Ostrogradsky theorem to transform
 \be
M = \sum_n\dfrac1{4\pi}\oint_{\Sigma_n} D^\nu k^\mu d\Sigma_{\mu\nu}
+ M_E, \lb{koM1}
 \ee
where $\Sigma_n$ denotes collectively all the spacelike boundary two-surfaces described above, and $M_E$ is the bulk integral
 \be\lb{ME}
M_E = \frac1{4\pi}\int D_\nu D^\nu k^\mu dS_\mu = -\frac1{4\pi}\int {R^\mu}_\nu k^\nu dS_\mu
= - 2\int {T^\mu}_\nu k^\nu dS_\mu,
 \ee
with
 \be
{T^\mu}_\nu = \dfrac1{4\pi}\left[F^{\mu\rho}F_{\nu\rho} -
\dfrac14\delta^\mu_\nu F^{\rho\sigma}F_{\rho\sigma}\right]
 \ee
the electromagnetic energy-momentum tensor. Applying again the three-dimensional Ostrogradsky theorem
to (\ref{ME}), we arrive at the decomposition of the Komar mass as the sum over rod contributions
 \be\lb{Mtot}
M = \sum_n M_n,
 \ee
where the masses of the constituents, including the corresponding bulk contributions, are expressed entirely in terms of the data on the axis (for more details see \cite{Clement:2017otx}):
 \be\lb{Mn}
M_n = \frac1{8\pi}\oint_{\Sigma_n}\left[g^{ij}g^{ta}\partial_j g_{ta}
+2(A_t F^{it}-A_\varphi F^{i\varphi})\right]d\Sigma_i.
 \ee
In this formula $x^a=t,\,\varphi$
and the remaining coordinates are labelled by $i,j$ (here not necessarily the Weyl coordinates $\rho,z$).
This formula, and a similar one for the Komar angular momentum, were not written down explicitly
by Tomimatsu, but are implicit in his derivation. Let us also note that Tomimatsu included in
the sum only the contributions of the various horizon components, whereas the sum should run over
all the cylinders surrounding the coordinate singularities -- horizons and strings -- on the polar axis.

For the total angular momentum, similar steps transform the Komar integral (\ref{koMJ}) into the sum of rod integrals
 \be\lb{Jtot}
J = \sum_n J_n, \qquad J_n =
-\frac1{16\pi}\oint_{\Sigma_n}\left[g^{ij}g^{ta}
\partial_jg_{\varphi a} +4A_\varphi F^{it}\right]d\Sigma_i.
 \ee
However, care should be taken in using (\ref{Jtot}) in the presence of Dirac or Misner strings extending to infinity, which is necessarily the case
if the total magnetic or NUT charge is non-zero. As previously noted in \cite{Clement:2017otx}, the electromagnetic contributions to (\ref{Jtot})
 \be
J_n^E = -\frac1{4\pi}\oint_{\Sigma_n}A_\varphi F^{it}d\Sigma_i
 \ee
transform under a large gauge transformation $A_\varphi \to A_\varphi + C$, such as used to shift the relative strength of Dirac strings, as
 \be
\Delta J_n^E = -\frac{C}{4\pi}\oint_{\Sigma_n}F^{it}d\Sigma_i = CQ_n ,
 \ee
where $Q_n$ is the electric charge carried by the rod $n$. The consequence is that, for a dyonic configuration with non-vanishing global electric and magnetic charges,
the sum (\ref{Jtot}) should also include an additional contribution from the surface at infinity
 \be
J_\infty^E = -\frac1{4\pi}\oint_{\Sigma_\infty}A_\varphi F^{it}d\Sigma_i,
 \ee
which vanishes only if the two Dirac strings extending to infinity are arranged symmetrically so that their contributions cancel each other. The situation is even worse in the presence of a global NUT charge. In this case the Komar integral itself
 \be\lb{koJ1}
J = \frac1{16\pi}\oint_{\Sigma_\infty}g^{ij}g^{ta}\partial_jg_{\varphi a} d\Sigma_i
 \ee
depends on the choice of gauge (on the relative strengths of the two Misner strings extending to infinity) and, as we shall see on the example of the Kerr-NUT solution, diverges unless the two strings are arranged symmetrically, corresponding to the choice $s=0$ in (\ref{s}).

\subsection{Revisiting the Tomimatsu representation}
Now we address the second logical step in the derivation of the Tomimatsu formulas. We use for the
rotating metric and electromagnetic one-form the standard axisymmetric Weyl-Papapetrou parametrization
 \ba\lb{weyl}
ds^2 &=& -F(dt-\omega d\varphi)^2 + F^{-1}[e^{2k}(d\rho^2+dz^2)+\rho^2d\varphi^2], \nn\\
A &=& vdt + A_\varphi d\varphi,
 \ea
and introduce the electromagnetic and gravitational
Ernst potentials,  which in Weyl  coordinates are defined by
 \be
{\cal E} = F - \overline{\psi}\psi + i \chi\,, \qquad \psi = v + iu\,,
 \ee
where the electric and magnetic scalar potentials $v$ and $u$ are
such that
 \be\lb{stat2}
v = A_t, \qquad \partial_i u =F \rho^{-1}\epsilon_{ij}\left(\partial_j A_\varphi +\omega  \partial_j v  \right)
\end{equation}
with $x^1=\rho$, $x^2=z$, and the twist potential $\chi$ is defined
by
 \be\lb{twist}
\partial_i\chi = -F^2\rho^{-1}\epsilon_{ij}\partial_j\omega
+ 2(u\partial_i v - v\partial_i u)\,.
\end{equation}
The Tomimatsu formulas  involve the imaginary parts of the electromagnetic and gravitational
Ernst potentials, $u = {\rm Im}\,\psi$ and $\chi = {\rm Im}\,{\cal E}$ \footnote{Some of our sign conventions differ from those used in \cite{tom84}.}. It is important that in view of the constancy of the rod directional vectors on each rod, the metric function $\omega$ takes a constant (and generally non-zero) value   $\omega_n$ along each rod, which defines the angular velocity of the corresponding object $\Omega_n=1/\omega_n$. Equally constant on each rod is the quantity
  \be
A_\varphi+\omega v = -\omega_n\Phi_n,
 \ee
defining the electric potential in the corotating frame
 \be\lb{phin}
-\Phi_n= A_t+\Omega_n A_\varphi.
 \ee
This generalizes the well-known property of the horizon $n=H$ \cite{carter}.

Now, combining gravitational and electromagnetic contributions to the rod mass
and passing to Ernst potentials we obtain in the limit $\rho\to0$, after integration over the cyclic coordinate $\varphi$
 \begin{align}\lb{Mn1}
M_n = &\frac1{4}\int_{z_n}^{z_{n+1}}\left[\omega\partial_z{\rm Im}\E +
2\partial_z(A_\varphi\,{\rm Im}\psi)\right]dz\nn\\=
&  \frac{\omega_n}{4}{\rm Im}\E\Big {\vert}^{z_{n+1}}_{z_n} +\frac12 (A_\varphi\,{\rm Im}\psi) \Big{\vert}^{z_{n+1}}_{z_n}
 \end{align}
(for more details, see \cite{Clement:2017otx}), which for $n=H$ differs from Tomimatsu's \cite{tom84} Eq. (52) by the presence
of the second term.  Similar transformations for the angular momentum (\ref{Jtot}) lead to
 \begin{align}\lb{Jn}
  J_n =&  \frac1{8}\int_{z_n}^{z_{n+1}}\omega\left[-2 + \omega\partial_z{\rm Im}\E + 2\partial_z(A_\varphi\,{\rm Im}\psi) - 2\omega
\Phi\partial_z{\rm Im}\psi\right]dz\nn\\=&
 \frac{\omega_n}{4}  \left\{ -(z_{n+1}-z_n)    + \left[\omega_n \left( {\rm Im}\E /2-\Phi_n {\rm Im}\psi\right)+
 A_\varphi\,{\rm Im}\psi \right] \Big {\vert}^{z_{n+1}}_{z_n}\right\},
 \end{align}
which for $n=H$ is in agreement with Eqs. (54)-(55) of \cite{tom84}.

One also needs a similar expression for the electric charge of a rod.
The electric field in (\ref{koMJ}) is related to the Ernst potentials by
 \be\lb{elernst}
F^{ti} = \frac{g^{ij}}{g_{tt}}F_{tj} -
\frac{g_{t\varphi}}{g_{tt}}F^{\varphi i} = e^{-2k}\left[\partial_i v
+ \frac{F\omega}\rho\epsilon_{ij}\partial_ju\right],
 \ee
leading on the axis $\rho=0$ (with account for $\sqrt{|g|}=
e^{-2k}\rho F^{-1}$) to
 \be\lb{Qn}
Q_n=\frac1{4\pi}\int_{\Sigma_n}\omega\partial_z{\rm Im}\psi\,dzd\varphi=\frac{\omega_n}2{\rm Im}\psi \Big {\vert}^{z_{n+1}}_{z_n}.
 \ee
This expression was found by Tomimatsu \cite{tom84} for the horizon charge,
but it equally holds for any rod. Using it, one can rewrite the angular momentum of a rod as
\be\lb{Jnq}
J_n=\frac{\omega_n}2\left(-\frac{z_{n+1}-z_n}2+ M_n-Q_n\Phi_n   \right).
\ee
This formula is applicable both to timelike rods (horizons) and spacelike ones (line singularities).

\subsection{Smarr formulas for the rod masses}
Consider first a timelike rod, corresponding to some horizon, say $[z_1, z_2]$.
Any horizon connected component corresponds to $N^2 \equiv \rho^2/g_{\varphi\varphi} =
0$ with
 \be
g_{\varphi\varphi} = F^{-1}\rho^2-F\omega^2 > 0,
 \ee
and thus generically $F<0$ (except in the special case of non-rotating horizons).
The mass and angular momentum of the horizon are related by (\ref{Jnq}), where $\omega_H=\Omega_H^{-1}$
is the inverse angular velocity of the horizon. The length of the horizon rod in the expression for the
angular momentum can be related to the Bekenstein entropy $S_H = {\cal{A}}_H/4$, where ${\cal{A}}_H$ is the horizon area
\be\lb{Ar}
{\cal{A}}_H=\oint d\varphi\int_{z_1}^{z_2} \sqrt{|g_{zz}g_{\varphi\varphi}|} dz =2\pi \int_{z_1}^{z_2}\sqrt{|e^{2k|}}|\omega| dz
\ee
(according to (\ref{weyl}), the product $F^{-1}e^{2k}$ is positive). On the other hand, the Hawking temperature is $T_H = \kappa_H/2\pi$,
where $\kappa_H=\sqrt{|e^{-2k}|}/|\omega_H|$ is the surface gravity in Weyl coordinates. So one has:
\be\lb{TS}
T_HS_H = \frac{\kappa_H}{8\pi}{\cal{A}}_H= \frac{z_2-z_1}4.
\ee
It follows that (\ref{Jnq}) is equivalent to the usual Smarr horizon mass formula \cite{smarr}
 \be\lb{smarr}
M_H = 2\Omega_H J_H + 2T_H S_H + \Phi_H Q_H,
 \ee
for each black hole constituent, including horizons carrying also magnetic charges and/or
gravimagnetic (NUT) charges. As could be expected, this local horizon formula does not (and should not)
contain information on such global artefacts associated with magnetic and gravimagnetic charges as Dirac and Misner strings.

Although the Komar-Tomimatsu relations between horizon observables were derived for rotating black holes ($\Omega_H \neq 0$),
they still can be used in the {\em limit} $\Omega_H\to 0$. In this limit, out of the three Tomimatsu relations giving $Q_H$,
$M_H$ and $J_H$, only one survives, the Smarr relation for static black holes:
\begin{equation}
  M_H = 2T_HS + \Phi_HQ_H.
\end{equation}

For finite spacelike rods we can proceed similarly, except that $g_{\varphi\varphi}$ will now be negative in the case of Misner strings. Defining the rod angular velocity  $\Omega_n=1/\omega_n$, we obtain
\be\lb{ssmarr}
M_n=2\Omega_n J_n+\frac12 L_n +\Phi_n Q_n,
\ee
where $L_n=z_{n+1}-z_n$ is the rod length, $J_n$ and $Q_n$ are the rod angular momentum and charge. Note that this term can also be presented as the product
of a spacelike surface gravity ${\kappa_n}$ (\ref{kadef}) with the two-dimensional area  ${\cal{A}}_n$ of the defect.
Indeed, since on the $n$-th rod $\lim_{\rho\to 0}|g_{\varphi\varphi}|\to F_n\omega_n^2 \neq 0$,  the apparently one-dimensional ``line'' singularity has a finite {\em two-dimensional} area:
\be\lb{Ar1}
{\cal{A}}_n=\oint d\varphi\int_{z_n}^{z_{n+1}} \sqrt{|g_{zz}g_{\varphi\varphi}|} dz = 2\pi \int_{z_n}^{z_{n+1}}\vert e^k\omega \vert dz.
\ee
So, similarly with (\ref{TS}), one can write:
\be
 \frac{z_{n+1}-z_n}4 =\frac{\kappa_n}{8\pi}{\cal{A}}_n.
\ee
The analogy with (\ref{TS}) suggests that Misner strings could be assigned an entropy equal to one-fourth of their area, in line with past  \cite{Hunter:1998qe,Hawking:1998jf,Carlip:1999cy,Mann:1999pc,Mann:1999bt} and more recent \cite{Kubiznak:2019yiu,Ballon:2019uha,Bordo:2019tyh,Bordo:2019rhu} proposals. However we feel that such an interpretation deserves further investigations.

For infinite spacelike rods (e.g., Misner strings in Kerr-NUT) the length terms will give infinite rod angular momenta. As already mentioned, the global Komar angular momentum will be finite only for the symmetrical choice $s=0$ in (\ref{s}), such that the North and South string length contributions to the total angular momentum cancel out.

For stationary axisymmetric electrovacuum fields one has three conserved quantities: the mass, the angular momentum and the electric charges, which are given by Komar and Maxwell surface integrals. Their total values $M,\,J,\,Q$, computed at infinity, were reduced to the sum of contributions over the rods (\ref{Mtot}), (\ref{Jtot}), and (\ref{Qtot}). While the charges and area of each individual rod are related by a Smarr formula of the same form (\ref{smarr}) or (\ref{ssmarr}), there is no reason to expect that the global quantities will generically satisfy a relation of the same kind, as the various components of a multi-black hole system, horizons and strings, will in general have different masses, angular momenta, electric charges and surface areas. The best which can be done is to add together the various Smarr formulas, leading to the global Smarr formula for the total mass (\ref{koM1}) as a sum over horizon and string components
 \be\lb{gsmarr}
M = \sum_{H_n}\left(2\Omega_{H_n}J_{H_n} + \frac{\kappa_{H_n}{\cal A}_{H_n}}{4\pi} + \Phi_{H_n}Q_{H_n}\right) +
\sum_{S_n}\left(2\Omega_{S_n}J_{S_n} + \frac{\kappa_{S_n}{\cal A}_{S_n}}{4\pi} + \Phi_{S_n}Q_{S_n}\right) .
 \ee
We now discuss briefly two examples where this global Smarr formula can be applied.

 \subsection{Kerr-NUT}

In \cite{Bordo:2019rhu} the Kerr-NUT metric (\ref{s}) was considered with asymmetric Misner strings if $s\neq 0$. In this case, the metric function $\omega$ behaves at infinity as
 \be
\omega \sim -2(n\cos\theta + s),
 \ee
so that the Komar angular momentum (\ref{koJ1}) evaluated on a large sphere of radius $R$ has a divergent contribution
 \be
J \sim \frac{R}4\int_0^\pi\omega\sin\theta\,d\theta \sim sR.
 \ee
This divergence was pointed out in \cite{Bordo:2019rhu}, where it was suggested to cancel it by defining
the ``total angular momentum'' as (in our notation) $J_{\rm tot} \equiv J - J_+ - J_-$,
which from (\ref{Jtot}) is nothing but the horizon angular momentum $J_H$. Indeed from (\ref{Jnq}), the two Misner strings extending up to $\pm R$ have divergent angular momenta
 \be
J_\pm \sim - \frac{R\omega_\pm}4 = \frac{s\pm n}2R,
 \ee
which add up to $sR$. The  physical total angular momentum $J$ can be finite only for the choice $s=0$, leading
to a symmetrical Misner string configuration (as previously noted in \cite{manko2005}), which we shall
now assume.

Passing  in (\ref{s}) from Weyl coordinates $\rho, z$ to prolate spheroidal coordinates
\begin{align}
&\rho=\sigma(x^2-1)^{1/2}(1-y^2)^{1/2},\quad z=\sigma xy,\\
&ds^2 = -F(dt-\omega d\varphi)^2 + F^{-1}\left[e^{2k}\sigma^2(x^2-y^2)\left(\frac{dx^2}{x^2-1} +
\frac{dy^2}{1-y^2}\right)+\rho^2d\varphi^2\right],
 \end{align}
with $\sigma^2=m^2+n^2-a^2$, one finds for $s=0$
 \begin{align}\lb{KNUT}
&F = \frac{f}\Sigma, \quad  e^{2k} = \frac{f}{\sigma^2(x^2-y^2)},\quad  \Sigma = (\sigma
x+m)^2 + (ay+n)^2,\quad  f = \sigma^2(x^2-1) - a^2(1-y^2),
\nn\\&
 \omega  f= - 2yn\sigma^2 (x^2-1)- 2a(1-y^2)(m\sigma x+m^2+n^2 ) .
 \end{align}
The event horizon rod is $x=1, y\in[-1,\,1]$, with the parameters $\kappa_H$, $\omega_H=1/\Omega_H$ given in (\ref{parakn}).
The dualization equation gives the twist potential
 \be\lb{chiKNUT}
\chi={\rm Im}\E = 2 \Sigma^{-1} (may-n\sigma x),
 \ee
with the values at the turning points $x=1, y=\pm1$:
 \be
 \chi_{\pm}=-\frac{n(n\mp a)}{2r_H}
 \ee
($r_H=m+\sigma$). Using the above formulas it is easy to find:
 \begin{align}
& M_H=\frac{\omega_H}{4}\left( \chi_+-\chi_-\right)=\frac{a\omega_H}{2r_H}=m+\frac{n^2}{r_H}\\
&J_H=\frac{\omega_H}2\left(M_H-\frac{L_H}2\right)=\frac{a^2\omega_H}{2r_H}=aM_H,\\
 &M_{\pm}=\mp\omega_{\pm}\chi_{\pm}=-\frac{n(n\mp a)}{2 r_H},\\
&J_{\pm}=\frac{\omega_\pm}2\left( M_\pm-\frac{L_\pm}2\right) ,
 \end{align}
where $\omega_\pm=\mp2n$, $L_H=2\sigma$ and $L_\pm=R-\sigma$, with $R$ a regularization length of the infinite rods ($R\to \infty$).
For our symmetrical setting, $s=0$, the sum of the string angular momenta is finite:
 \be
 J_+ + J_-= -n(M_+ - M_-)=-\frac{an^2}{r_H}=a(M_+ + M_-).
 \ee
Note that the Kerr proportionality holds separately for the horizon rod, and for the sum of the strings, as well as for the global quantities
\be
J=J_H+ J_+ +J_-=a(M_H+M_+ + M_-)=aM,
\ee
where the total mass $M$ has the value $m$. Note also that the strings are always rotating in opposite directions,
$\Omega_\pm=\mp 1/(2n)$, even in the case where the horizon is non-rotating, $a=0$. But in this case the sum
of their angular momenta is zero.

The global Smarr relation (\ref{gsmarr}) for Kerr-NUT can be written in the form
 \be\lb{gsmarrkn}
M = 2T_HS_H + 2\Omega_HJ_H + 2\Omega_+\tJ_+ + 2\Omega_-\tJ_-
 \ee
where the ``reduced string angular momenta''
 \be
\tJ_\pm \equiv J_\pm + \frac{\omega_\pm L_\pm}4 = \frac{\omega_\pm M_\pm}2 = \pm\frac{n^2(n\mp a)}{2r_H}
 \ee
can be considered as the finite part of the upper/lower Misner string angular momentum.The Smarr relation (\ref{gsmarrkn})
is equivalent to the relation (10) of \cite{Bordo:2019rhu},  where $\Omega$ and $J$ should be understood as our $\Omega_H$ and $J_H$,
the ``Misner potentials'' are related to our string angular velocities by $\psi_\pm = \mp\Omega_\pm/4\pi$, and the ``Misner charges'' to
our reduced string angular momenta by $N_\pm = \mp4\pi\tJ_\pm$. It is interesting to note that, similarly to Eq. (9) of \cite{Bordo:2019rhu},
the differential Smarr relation, or generalized first law,
 \be
dM = T_HdS_H + \Omega_HdJ_H + \Omega_+d\tJ_+ + \Omega_-d\tJ_-,
 \ee
is satisfied by the Kerr-NUT solution.

\subsection{Dyonic Kerr-Newman-NUT}

To add magnetic and NUT charges to the electric Kerr-Newman solution in spheroidal coordinates, one has merely to complexify the Ernst potentials of the latter
$m\to m+in,\, q\to q-ip$ ($q$ and $p$ being the electric and magnetic charge parameters), leading to
 \be\lb{ernstKNN}
\E = \frac{\sigma x - m + i(ay-n)}{\sigma x + m + i(ay+n)}, \qquad \psi =
\frac{-q+ip}{\sigma x + m+ i(ay+n)},
 \ee
where now $\sigma^2=m^2+n^2-p^2-q^2-a^2$.
The resulting scalar electromagnetic potentials are
 \be
\Sigma v = - q(\sigma x+m) + p(ay+n) ,\qquad \Sigma u = p(\sigma x+m) + q(ay+n),
 \ee
where $\Sigma$ is the same as in (\ref{KNUT}), with the new $\sigma$, and the corresponding
twist potential is that of (\ref{chiKNUT}). The metric functions are unchanged from those of (\ref{KNUT}), except for
\be
\omega  f = - 2ny\sigma^2(x^2-1) - 2a(1-y^2)(m\sigma x+m^2+n^2-e^2/2),
\ee
with $e^2 = q^2+p^2$. Finally, the electromagnetic four-potential is
 \be
A = v\,dt + \left(-py +\left[2ny- a(1-y^2) \right]v\right)\,d\varphi,
 \ee
where, as advocated in \cite{Clement:2017otx}, we have set the gauge so that the two Dirac strings are symmetric.

The calculation of the electric charge, mass and angular momentum of the horizon rod, and of the conjugate variables, gives:
 \begin{align}
& Q_H=\frac{2(mq-np)r_H-q e^2}{\nu^2-4a^2n^2\nu^{-2}},\qquad \Phi_H= \frac{qr_H-np}{\nu^2},\qquad \Omega_H = \frac{a}{\nu^2},\\
&M_H=\frac{2(m^2+n^2) r_H-me^2}{\nu^2-4a^2n^2\nu^{-2}}+\left( pe^2/2-\mu r_H\right)\left[ \frac{pr_H+q(n+a)}{(\nu^2+2an)^2} + \frac{pr_H+q(n-a)}{(\nu^2-2an)^2}\right],\\
&J_H=\frac{\omega_H}2\left( -\sigma+M_H- \Phi_H Q_H \right),
 \end{align}
where we have put $\nu^2=r_H^2+n^2+a^2 = 2(mr_H+n^2-e^2/2)$, and $\mu=pm+qn$.
Similarly, for the North/South combined Misner and Dirac strings labeled by $\pm$ we find
\begin{align}
& Q_\pm = \frac{n(pr_H+qn_\pm)}{\nu^2\pm 2an},\qquad \Phi_\pm= -\frac{p}{2n},\qquad \Omega_\pm = \mp\frac1{2n},\\
& M_\pm = -\frac{n(nr_H - mn_\pm)}{\nu^2\pm 2an} + \frac{\left(\mu r_H-pe^2/2 \right)\left(pr_H+qn_\pm\right)}{(\nu^2\pm 2an)^2},\\
& \tJ_\pm = \mp n \left[\frac{(p^2-n^2)r_H + (mn+pq)n_\pm}{\nu^2\pm 2an} + \frac{n\left(qr_H-pn_\pm \right)\left(pr_H+qn_\pm\right)}{(\nu^2\pm 2an)^2}\right],
\end{align}
with $n_\pm = n \pm a$. Together these satisfy the global Smarr relation
 \be\lb{gsmarrknn}
M = 2T_HS_H + 2\Omega_HJ_H + \Phi_HQ_H + 2\Omega_+\tJ_+ + 2\Omega_-\tJ_- + \Phi_+Q_+ + \Phi_-Q_-,
 \ee

Although the Tomimatsu approach for the strings breaks down for $n=0$ because $\Omega_\pm$ diverges, we can nevertheless recover the results of \cite{Clement:2017otx} for the dyonic Kerr-Newman black hole by taking with due care the {\em limit} $n\to0$. In this limit the string electric charges $Q_\pm = nu_\pm$ go to zero, so that $Q_H=Q$, but the string potentials $\Phi_\pm = -p/2n$ diverge, their product going to the finite limit $-(p/2)u_\pm$. Likewise, the two reduced string angular momenta $\tJ_\pm$ go to zero, so that $J_H=J$. However the string angular velocities $\Omega_\pm$ diverge, so that their product goes to a finite limit
 \be
\Omega_\pm\tJ_\pm = \frac12(M_\pm - Q_\pm\Phi) \to pu_\pm,
 \ee
where we have used (\ref{Mn}) to compute the limit $M\pm \to (p/2)u_\pm$. The string area also goes to zero, so that the global Smarr relation (\ref{gsmarrknn}) reduces to the form
  \be\lb{gsmarrknd}
M = 2T_HS + 2\Omega_HJ + \Phi_HQ + \Psi_HP,
 \ee
where $P=p$ is the magnetic charge, and $\Psi_H= p(u_++u_-)/2 = pr_H/\nu^2$ can be interpreted as  an effective  horizon magnetic potential.

 \setcounter{equation}{0}
\section{Summary and outlook}
We expanded and reformulated Tomimatsu's representation of ​​Komar  charges in a stationary axisymmetric asymptotically locally flat space-time containing black holes and weak string-like defects on the symmetry axis. In Weyl coordinates, both black holes and defects can be described universally as rods located on the  axis, the difference between the horizons and defects being in the signature of their directional vectors. Using the three-dimensional Ostrogradsky theorem, one can express the global charges as a sum of rod contributions, paving the way to obtain mass formulas in the presence of defects. Constancy of the above vectors along the defects allows to define their angular velocities. In spacetime, directional vectors define the corotating Killing vectors which become null both on the black hole horizons and the defects, which allows to define analogue surface gravity for the latter.

We found explicit algebraic formulas for rod charges in terms of Ernst potentials at turning points.  The resulting Smarr formula for the horizon mass does not include  a magnetic charge or NUT contribution (correcting numerous errors in the literature).
The length term, present in the Tomimatsu angular momentum for the horizon rods has a dual interpretation as an entropy term. Similar terms for defects also admit a dual presentation as the product of the analogue surface gravity of the corotating Killing vector with the defect area. However, we refrain from giving to this term a true entropy meaning since the corotating Killing vector in the vicinity of a defect is spacelike, so the associated Killing horizon presents a situation different from that of both external and internal black hole horizons.

Having applied this formalism to the Kerr-NUT spacetime, we found that finiteness of the total angular momentum  selects the symmetric Misner string gauge, in which case  the Kerr rule holds both for the total angular momentum,  $J=aM$  and for the sum of angular momenta of the Misner  strings. The total mass is the sum of the Komar mass computed on the horizon and the sum of the masses of Misner strings, the same balance holds for angular momenta.  The Smarr relation which we obtained for the total mass includes the sum of the products of the angular momenta of the two Misner strings with their angular velocities.

In the case of the Kerr-Newmann-NUT solutions with a magnetic charge, the expressions for the  individual string masses and angular momenta  look more complicated and exhibit a complex non-linear character due to the superposition of Misner and Dirac strings. The total mass formula  now also involves the products of the electric charges of the two Dirac strings with their electric potentials. In the limit of a vanishing NUT charge, the string terms in this formula reduce to an effective horizon magnetic potential.

In an Appendix we show (extending the previous result for $s=0$) that for  Kerr-NUT instantons with an arbitrary string tuning parameter $s$,  one can not ensure both the Misner and Hawking periodicities globally, so that the Euclidean solutions are always plagued with orbifold singularities.   Therefore, one can not extend the standard calculations of the instanton actions with NUTs to the rotating case.   Still, this does not mean that the resulting action will be infinite.

Our new findings further support the proposal of physical interpretation of NUTty black holes without imposing the Misner periodicity condition on time.
Combining the Tomimatsu approach with the rod structure we obtained a convenient framework  which may be essentially useful for the analysis of multi-center solutions containing several horizon components and cosmic string, Misner string and Dirac string line singularities. The resulting global Smarr-type mass formulas generically will not have a form as simple as for single black holes, but will involve contributions of the various horizon and string angular momenta and areas. The physical interpretation of the string area terms as entropy does not seem clear at this moment, but we expect to come back to this question shortly.

While this work was being finalized, there appeared a preprint \cite{durka} focussing on the contribution of the Misner string angular momenta and area to black hole thermodynamics, in the special case of non-rotating Taub-NUT spacetime.

\section*{Acknowledgments} DG thanks LAPTh Annecy-le-Vieux for
hospitality at different stages of this work. He also acknowledges
the support of the Russian Foundation of Fundamental Research under
the project 17-02-01299a and the Russian Government Program of
Competitive Growth of the Kazan Federal University.

\renewcommand{\theequation}{A.\arabic{equation}}
\section*{Appendix. Non-existence of regular Kerr-NUT instanton with asymmetric Misner strings }
The Euclidean version of the Kerr-NUT metric with arbitrary gauge parameter $s$ reads
\begin{align}\label{ekn}
 ds^2 &=
    \frac{\Delta}{\Sigma}( dt + P_\theta d\varphi)^2 +  \Sigma \left(\frac{dr^2}{\Delta} + d\theta^2\right) + \frac{\sin^2\theta}{\Sigma} (P_r d\varphi + a dt)^2,\;\; \Sigma  = r^2 - (n + a\cos\theta)^2,
    \\
 P_r  &= r^2 - n^2 - a^2 + 2sa,\;\;
 P_\theta  = 2n\cos\theta - a\sin^2\theta +2s,  \;\;
 \Delta = r^2 - 2mr + n^2 - a^2.
 \end{align}
The normalized rod directions are given by (\ref{lh},\, \ref{ln}) with indices $n=\pm, H=B$ (bolt, where $\Delta$ vanishes) and $r_B=m+\sqrt{m^2+a^2-n^2}$:
\be
 \kappa_B=\frac{r_B-m}{(mr_B-n^2+as)},\quad \Omega_B=\frac{-a \kappa_B}{2(r_B-m)}, \quad \kappa_{\pm}=\mp\frac1{2(n\pm s)},\quad\Omega_{\pm}= \kappa_\pm.
\ee
Now all the three direction vectors are spacelike, and to avoid conical and orbifold singularities one has
to identify both the Euclidean time $t_E$ and $\varphi$ in accordance with (\ref{ident}). Suppose we have done
this at the South pole, where meet the rods $L_-,\, l_B$. Then, at the North pole generically we have to do
another identification of $t_E$ and $\varphi$ to match the pair $ l_B,\,l_+$. At least, both identifications
have to be compatible up to multiplication on some integers. In other words, if one introduces $2\times 2$
matrices,  $W_S,\, W_N$, formed by the rows of the pairs $W_S=(l_-,\, l_B)$ and $W_N=( l_B,\, l_+)$, one must
ensure that $W_S\cdot W_N^{-1}\in GL(2,Z)$. It is easy to compute that
\be
g=W_S\cdot W_N^{-1}=\begin{pmatrix}
  \alpha& \beta\\
  1& 0
\end{pmatrix},\quad \alpha=\frac{4n\kappa_B}{1+2(n+s)\Omega_B} ,\quad
\beta =\frac{1+2(s-n)\Omega_B} {1+2(n+s)\Omega_B}.
\ee
The necessary condition of matching is $\det g =\pm 1$, implying $\beta=\pm 1$. For the upper sign one has $n=0$ or $a=0$, i.e. Kerr-bolt or Euclidean Taub-NUT (known).
For the lower sign one obtains the equation
\be
1+2s\Omega_B=0.
\ee
The solution reads
\be
m^2=\frac{n^4}{n^2+a^2}.
\ee
But in this case, the metric function $\Sigma$ is not positive definite on the bolt, violating the metric signature unless $a=0$.

Finally, there is the possibility that the vectors $(l_-,\, l_B)$ are parallel (or  the pair $( l_B,\,l_+$). Then we have only two rods, and  no problem of matching arises.
But from these conditions one finds $m=n=s$. It turns out that the solution (\ref{ekn}) can then be reduced to the Taub-NUT instanton  by some coordinate transformation \footnote{These results were obtained in collaboration with Dmitry Torbunov.}.

\end{document}